\documentclass[twocolumn,pra,aps]{revtex4}
\usepackage{graphicx}

\begin{document}

\title{The General Free Will Theorem}

\author{Antoine Suarez}
\address{Center for Quantum Philosophy, P.O. Box 304, CH-8044 Zurich, Switzerland\\
suarez@leman.ch, www.quantumphil.org}

\date{June 12, 2010}

\begin{abstract} It is argued that the Strong Free Will Theorem does not prove \emph{nonlocal determinism} wrong. This is done by the before-before (Suarez-Scarani) experiment, which is used here to prove the following \emph{General Free Will Theorem}: If humans have a certain amount of free will, there are other free beings outside space-time producing nonlocal effects in our world, which are both random and lawful.\\

\end{abstract}


\maketitle

The proof of the Free Will Theorem (FWT) presented by John Conway and Simon Kochen in 2006 uses the Kochen and Specker theorem \cite{sfwt}. It is however convenient to formulate it in the context of the Bell theorem \cite{jb} in order to relate  the FWT to experiments that have already been done. In this context the FWT amounts to stating that the assumption of determinism and locality are incompatible with the predictions of quantum mechanics in entanglement experiments, provided the experimenters are free in choosing certain relevant parameters at each side of the setup. This is much the same as Bell theorem states, and so far the FWT bestows upon us nothing substantially new \cite{tu,ng}, other than explaining the deep philosophical meaning of Bell's discovery: if humans have free will, there is free will behind the quantum phenomena. Undoubtedly this is not of little merit.

The FWT proves only (\emph{local determinism}) wrong, i.e., the assumption that the outcomes of an experiment are determined by the information accessible to the particles from their \emph{past light cones} (time-ordered local causality). With the ``Strong FWT''in 2009 the authors aimed to strengthen the theorem and refute \emph{nonlocal determinism} as well, by allowing the outcomes to depend also on information coming from \emph{past half-spaces} (time-ordered nonlocal causality) \cite{sfwt}.

This Letter shows that the ``Strong FWT'' does not consistently refute \emph{nonlocal determinism}. This is rather done  by the before-before (Suarez-Scarani) experiment \cite{asvs97, as00.1, szsg}, which is used in the follow to generalize the FWT.

\begin{figure}[t]
\includegraphics[width=80 mm]{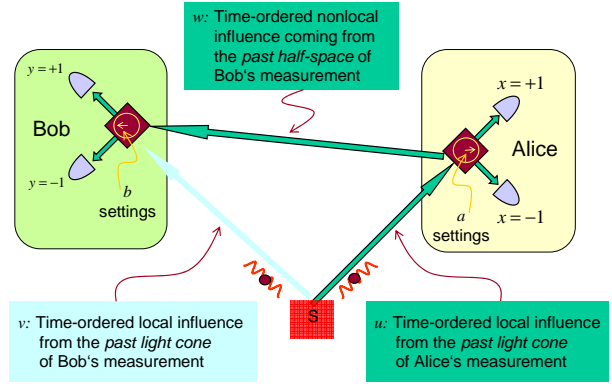}
\caption{Suarez-Scarai model: Full deterministic causal chain (dark-green) built of time-ordered nonlocal influences \emph{w} and time-ordered local ones \emph{u,v}. The individual outcomes \emph{y} depends nonlocally on pre-existing \emph{x} outcomes, and these depend locally on the variable \emph{u}; a similar (not shown) causal chain could be drawn (light-blue, clockwise) for the alternative case of the individual \emph{x} outcomes depending nonlocally on \emph{y} outcomes. The model is non-sigaling (see text).}
\label{f1}
\end{figure}

We start by considering standard Bell experiments like the one sketched in Figure \ref{f1}. A source emits pairs of photons in a entangled state. One of the photons is sent to Alice's laboratory and measured with a convenient device (usually an interferometer or a polarizing beam-splitter), and the other photon is sent to Bob's laboratory and measured with a similar device. Alice sets the parameter \emph{a} (likewise Bob for \emph{b}) and gets the outcome \emph{x} (respectively \emph{y}). Alice's and Bob's measurements happen spacelike separated from each other.

Classically, the value \emph{x} is a function of the setting \emph{a} and the variable \emph{u} representing the information accessible from the \emph{past light cone} at the moment of Alice's measurement:
\begin{eqnarray}\label{1}
  x=F(a,u)
\end{eqnarray}

and similarly for Bob's measurement:
\begin{eqnarray}\label{2}
  y=G(b,v)
\end{eqnarray}

John Bell proved that the outcomes described by the Equations (\ref{1}) and (\ref{2}) fulfill certain locality criteria called Bell inequalities \cite{jb}. Bell experiments demonstrate violation of such inequalities. Therefore it holds:
\begin{small}
\begin{eqnarray}\label{3}
  &&\texttt{Experimenter's freedom + Bell experiment} \nonumber\\
  &&\texttt{+ Time-ordered local causality} \nonumber\\ &&\Rightarrow \texttt{contradiction}
\end{eqnarray}
\end{small}
Where the axiom \emph{Experimenter's freedom} means: The choices Alice and Bob make (\emph{a}, \emph{b}) are uncorrelated to the properties the particles may have in each run (``there is no conspiracy''). In more technical terms: the choices $a$ of Alice and the responses $y$ of Bob's particle (the particle measured by Bob), are not determined by information contained in the common region of the past light cones of Alice and Bob's particle. And likewise for the choices $b$ of Bob and the responses $x$ of Alice's particle.

Equation (\ref{3}) implies that the outcomes of a Bell experiment are not determined by the information accessible from the \emph{past light cone} of each measurement. Conway and Kochen argue that this is much the same as stating that the particles produce their responses (\emph{x} and \emph{y}) as freely as the experimenters Alice and Bob choose their settings (\emph{a} and \emph{b}). Thus from (\ref{3}) one is led to the FWT:
\begin{small}
\begin{eqnarray}\label{4}
  \texttt{Experimenter's freedom} \Rightarrow \texttt{Particle's freedom}
\end{eqnarray}
\end{small}
Theorem (\ref{4}) is nothing other than Bell's theorem (\ref{3}) expressed in philosophical terms.

Consider now the situation in which all apparatuses are at rest in the laboratory frame, and in this frame Alice makes her measurements a bit before Bob does, like represented in Figure \ref{f1}. If one admits \emph{time-ordered nonlocal causality}, Alice's outcome happening before in time can be considered to cause Bob's outcome happening later in time. Then it is possible to built a full deterministic causal chain consisting of time-ordered nonlocal influences (\emph{w}) and time-ordered local ones (\emph{u}) such that all outcomes are determined by information accessible from the \emph{past}(half-space) at the time of measurement.

If one permits nonlocal dependencies like $w$, then one has to reformulate more precisely the axiom of \emph{Freedom}: To say that Alice's choice of $a$ (respectively Bob's choice of $b$) is free means ``that it is not determined by (i.e., is not a function of) what has happened at earlier
times (in any inertial frame).'' \cite{sfwt}

Accordingly the ``Strong FWT'' states that the response of Alice's particle (respectively Bob's one) is free in exactly the same sense, i.e., ``it is not a function of properties of that part of the universe that is earlier than this response with respect to any given inertial frame.'' \cite{sfwt}. In other words, the particle's responses cannot be explained by information accessible from the \emph{past}(half-space) (in any frame).

The proof depends decisively on the following claim the authors call ``Axiom MIN'':

MIN: Assume that the experiments performed by Alice and Bob are space-like separated. Then Bob can freely choose his setting $b$, and Alice's outcome $x$ (i.e. the response of the photon Alice measures) is independent of Bob's choice. Similarly and independently, Alice can freely choose her setting $a$ and Bob's outcome $y$  (i.e. the response of the photon Bob measures) is independent of Alice's choice. \cite{sfwt}

With MIN one excludes all possible time-ordered nonlocal influences like $w$ in Figure \ref{f1}. Thus, if a particles' response is determined by properties of its past, these are properties of the particle's past light cone. But this is ruled out by Theorem (\ref{3}). Thus the ``Strong FWT'' follows straightforwardly from MIN.

Conway and Kochen assert that MIN is motivated by relativity and hence their result is physically relevant: Freedom is an attribute of particles in the real universe.

However, it is possible to deny MIN without conflicting with any experiment or falling into absurdities, unless one takes into consideration the before-before experiment \cite{szsg}, and therefore Conway and Kochen's proof is not consistent. I show this in the following:

These authors argue that influences like \emph{w} are at odds with relativity because in a different frame Alice may measure after Bob, and thus Bob's measurement cannot be influenced by Alice's one \cite{sfwt}.

One can formulate this argument more accurately \cite{ng}:

Suppose that in a frame where Alice measures before Bob her outcome \emph{x} is given by the function (\ref{1}), and in a frame when she measures after Bob by the function:
\begin{eqnarray}\label{5}
  x=F'(a,u,b,v,w)
\end{eqnarray}

And likewise for Bob:
\begin{eqnarray}\label{6}
  y=G'(b,v,a,u,w)
\end{eqnarray}
If one assumes that the outcomes are frame invariant in any case, then from (\ref{1}) and (\ref{5}) it follows:
\begin{eqnarray}\label{7}
F(a,u)=F'(a,u,b,v,w)
\end{eqnarray}
Equation (\ref{7}) implies that the function $F'$ does not depend on the variables \emph{b}, \emph{v} and \emph{w}, and therefore it is necessarily local. Similarly one proves that the function $G'$ is local as well. And according to (\ref{3}) local functions cannot account for the correlations in Bell experiments.

Nonetheless, the preceding argument does not exclude time-ordered nonlocal causality: That such a causality conflicts with ``covariance'' does not necessarily mean that it conflicts with experiment or bears absurdities.

\begin{figure}[t]
\includegraphics[width=80 mm]{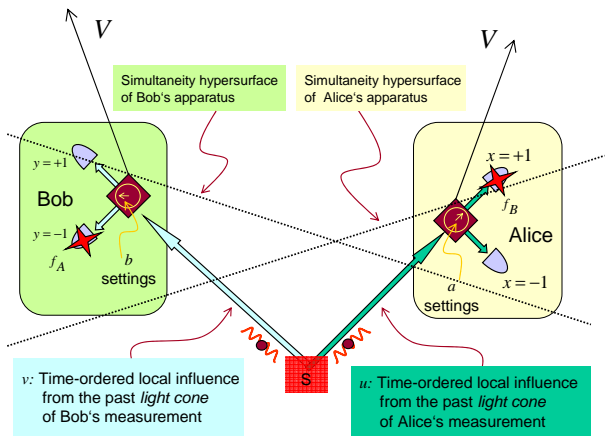}
\caption{Before-before experiment: In the inertial frame of Alice's apparatus her measurement happens before Bob's one.
In the inertial frame of Bob's apparatus his
measurement happens before Alice's one. The nonlocal correlations don't disappear: The experiment refutes nonlocal determinism.}
\label{f2}
\end{figure}

Effectively Conway and Kochen seek to complete their argument by stating: ``To accept relativity but deny MIN is therefore to suppose that an experimenter can freely make a choice that will alter the past, by changing the location on a screen of a spot that has already been observed \cite{sfwt}''. In other words, Conway and Kochen claim that time-ordered nonlocal influences necessarily imply \emph{signaling faster than light}, and in their proof exclude frame-dependent probability distributions in principle (Reference \cite{core} excludes them too). However it is possible to construct testable non-signaling models that assume probability distributions depending on the state of movement of the measuring devices:

Consider the before-before experiment sketched in Figure \ref{f2}. By setting the measuring devices in movement one  gets different timings. The Suarez-Scarani model \cite{asvs97, as00.1, as09} states that if Alice's apparatus is first to select the outcome in its inertial frame, then the outcome \emph{x} is given by a function depending on local variables like in Equation (\ref{1}). However, if Alice's apparatus selects after in its inertial frame, \emph{x} is given by a function depending on the settings \emph{a, b} and the nonlocal variable \emph{w}:
\begin{eqnarray}\label{8}
  x=F''(a,b,w)
\end{eqnarray}

Similarly, if Bob's apparatus selects before in its inertial frame, the outcome \emph{y} is given by Equation (\ref{2}), and if it selects after, \emph{y} is given by:
\begin{eqnarray}\label{9}
  y=G''(b,a,w)
\end{eqnarray}
Where the dependencies expressed by $F''$ and $G''$ are obviously supposed to reproduce the quantum mechanical correlations.

To date there is no proof that correlations defined according to (\ref{1}), (\ref{2}), (\ref{8}) and (\ref{9}) are signaling \cite{vang}, and I dare to conjecture that the quantum mechanical formalism prevents such a proof.

The Suarez-Scarani model combines local and nonlocal hidden variables to a full time-ordered nonlocal causal explanation of the quantum correlations, which remains compatible with both Bell- and  Michelson-Morley-type experiments and is non-signaling. While the model can be considered relativistic, because it accepts multisimultaneity, it is not Lorentzinvariant, since it assumes distributions depending on the time-order. It illustrates that Lorentzinvariance it is not a necessary condition for non-signaling.

In before-before experiments the model makes predictions conflicting with quantum mechanics: If each measuring device, in its own reference frame, is first to select the output  as represented in Figure \ref{f2}, (\ref{8}) and (\ref{9}) become irrelevant and only (\ref{1}) and (\ref{2}) matter. Hence the model predicts disappearance of nonlocal correlations with maintenance of possible local ones. By contrast quantum mechanics predicts timing independent nonlocal correlations \cite{as09}.

Before-before experiments have been done and falsify \emph{nonlocal determinism} \cite{szsg}. Hence one has:
\begin{small}
\begin{eqnarray}\label{10}
  &&\texttt{Experimenter's freedom} \nonumber\\
  &&\texttt{+ Suarez-Scarani experiment} \nonumber\\
  &&\texttt{+ Time-ordered nonlocal causality} \nonumber\\ &&\Rightarrow \texttt{contradiction}
\end{eqnarray}
\end{small}
which completes Bell's theorem (\ref{3}).

In summary, it is not true in general that``the FWT shows that nature itself is non-deterministic'' \cite{sfwt}. This is rather shown by theorem (\ref{10}).

Notice that the deterministic causal chain of Figure \ref{f1} can have trivial local parts $u$ and $v$, and therefore is neither refuted by Leggett-type experiments \cite{as09} nor by the proposed Colbeck-Renner ones \cite{core}. By contrast Legget and Colbeck-Renner extensions of quantum mechanics, as far as they assume time-ordered nonlocal influences, are falsified by the before-before experiment \cite{as09}.

From (\ref{10}) I derive the \emph{General Free Will Theorem}:

If ``correlations cry out for explanation'' (as John Bell liked to say \cite{jb}), then the first implication of (\ref{10}) is that the ``explanation'' lies outside space-time: Nonlocal correlations cannot be explained by any narrative in space-time, ``the space-time does not content the whole physical reality'' (Nicolas Gisin).

Therefore expressions like ``free decisions of particles'' are confused since ``particle'' suggests something material located in space-time. It is more appropriate to say that the free decisions involved in the quantum effects come from beings outside space-time. I propose to call such beings \emph{'johnbells'}, in honor of the discoverer of nonlocality. Although \emph{johnbells} produce measurable effects, they themselves are unobservable. Following Conway and Kochen one can conclude: if experimenters have a certain freedom, then \emph{johnbells} have exactly the same kind of freedom \cite{sfwt}.

Additionally, theorem (\ref{10}) implies that \emph{johnbells} produce nonlocal effects in space-time (correlated separated events), which are both random and lawful: Randomness in Alice's lab and randomness in Bob's lab, but in both places \emph{the same} randomness: ``something very different from classical stochasticism is at play here''\cite{sfwt}. The fact that \emph{johnbells} do not produce mere randomness but controlled one \cite{as08.1} strengthens the conclusion that they have the same kind of freedom as humans.

Then it follows:
\begin{small}
\begin{eqnarray}\label{11}
  &&\texttt{Free human beings} \nonumber\\
  &&\Rightarrow \texttt{Free beings outside space-time}
\end{eqnarray}
\end{small}
This is the General Free Will Theorem, which generalizes theorem (\ref{4}).

It is interesting to connect this result with the ``Flash ontology'' (FO) proposed by Roderich Tumulka and coworkers (usually denoted rGWRf: ``relativistic GRW model with the flash ontology'') \cite{tu}. Here the quantum stuff are the locations (flashes) of the spontaneous localization. In the experiment of Figure \ref{f2} a pair of flashes $f_A, f_B$ happens when one of Alice's detectors and one of Bob's detectors jointly fire.

Conway and Kochen (in the Appendix of their 2009 paper) argue that the FWT refutes FO \cite{sfwt}. Here I show that this is not the case:

FO shares three main features: it is stochastic, nonlocal and covariant \cite{tu}.

However FO \emph{stochasticism} is very different from the classical one: It is not so that each particle generates random flashes; it is the pair of particles that jointly produce \emph{correlated} random ''flashes``, which cannot be explained by any history in space-time.

Regarding \emph{covariance} FO states: ``The objective facts are where-when the flashes occur, and it is enough if a theory prescribes, as does rGRWf, their joint distribution in a Lorentzinvariant way.'' \cite{tu}

This means: FO does not assume that nature chooses the $f_B$ first, and $f_A$ afterwards, or the other way around: One flash is not the cause of the other. Additionally, the probability distributions of possible future flashes is time-order invariant. The laws governing rGRWf are covariant.

According to (\ref{10}) time-order nonlocal influences do not exist. Thus each single pair of flashes ($f_B$,$f_A$) can be considered a single event the concept of covariance does not apply to. And there where this concept makes sense, i.e., the evolution of the probability distributions, FO is covariant.

Hence FO is perfectly compatible with the ``General FWT''.

Nonetheless, assuming that correlated events arise without time-ordered causality, and hence physical reality emerges from outside space-time, does not well fit to ``relativity''. The 'r' in 'rGRWf' is ambiguous and I would advice to avoid this shortcut using 'FO' instead.

The Suarez-Scarani model can be considered relativistic but not Lorentzinvariant, and is refuted by the before-before experiment. By contrast the FO is Lorentzinvariant but not relativistic, and is not refuted by the before-before experiment. Conway and Kochen mistakenly assume that the flashes of a pair ($f_B$,$f_A$) in FO are time-ordered, and time-ordered nonlocal influences necessarily imply signaling.

Regarding \emph{ontology}, FO corrects the wrong association of the wavefunction to a sort of ``cloud'' evolving in space-time. In this sense FO provides a better ontology than Copenhagen's collapse of the wavefunction.

However, the collapse picture has the advantage of showing that: 1) The concept of ``particle'' makes sense only at detection, and 2) Nonlocality is not necessarily coupled to Bell inequalities or other related locality criteria (GHZ), but does already happen in single particle experiments between the two detectors.

In conclusion: No experiment can prove ``the free will of the experimenter''. Everyone is free to choose the axiom of ``Determinism'' or the axiom of ``Freedom''. But if one chooses the latter, then the before-before experiment demonstrates the existence of free beings outside space-time (\emph{johnbells}) controlling the random behavior of the particles in our world.

Conway and Kochen assert: ``In the present state of knowledge, it is certainly beyond our capabilities to understand the connection between the free decisions of particles and humans, but the free will of neither of these is accounted for by mere randomness.''\cite{sfwt}

The \emph{General Free Will Theorem} suggests: An important step towards understanding the ``connection'' may be to say that the ``free decisions'' come from beings outside space-time rather than from ``particles'' in space-time.

Recently Stephen Hawking suggested that intelligent extraterrestrials are almost certain to exist — but that humanity should be doing all it can to avoid any contact.
In a comment it was joked that a clear sign for the intelligence of aliens is the fact that they seem to be doing all they can to avoid any contact with the human species. The General Free Will Theorem proved above rather suggests that intelligent beings dwelling outside space-time are uninterruptedly producing phenomena on Earth, and all over the Universe. The fact that we do not easily realize the presence of these invisible \emph{johnbells} is probably more a sign of mental weakness on our part than of reluctance to communicate with us on their part.\\

\noindent\emph{Acknowledgments}: I am grateful to Nicolas Gisin  and Roderich Tumulka for stimulating comments.


\begin{references}


\bibitem{sfwt} Conway J.H. and Kochen S., The Free Will Theorem. \emph{Found. Phys.} \textbf{36}, 1441-1473 (2006); quant-ph/0604079. The Strong Free Will Theorem. \emph{Notices of the American Mathematical Society} \textbf{56}, 226–232 (2009).

\bibitem{jb} Bell J. S., {\em Speakable and unspeakable in quantum mechanics}, Cambridge: University Press, 1987.

\bibitem{tu} Tumulka, R., Comment on "The Free Will Theorem". \emph{Found. Phys.} 37, 186–197 (2007). Goldstein Sh., Tausk D. V., Tumulka R., Zanghi N.,What Does the Free Will Theorem Actually Prove? arXiv:0905.4641v1 (2009).

\bibitem{ng} Gisin N., The Free Will Theorem, Stochastic Quantum Dynamics and True Becoming in Relativistic Quantum Physics. arXiv:1002.1392v1 (2010).

\bibitem{asvs97} Suarez A. and Scarani V., Does entaglement depend on the timing of the impacts at the beam-splitters? {\em Phys. Lett. A}, {\bf 232}, 9 (1997).

\bibitem{as00.1} Suarez A., Quantum mechanics versus multisimultaneity in experiments with acousto-optic choice devices. {\em Phys. Lett. A}, {\bf 269}, 293 (2000).

\bibitem{szsg}  Stefanov A.,  Zbinden H.,  Gisin N., and  Suarez A., Quantum Correlations with Spacelike Separated Beam Splitters in Motion: Experimental Test of Multisimultaneity. {\em Phys. Rev. Lett.} \textbf{88} 120404 (2002) and \emph{Phys. Rev. A} \textbf{67}, 042115 (2003).

\bibitem{core} Colbeck, R. and Renner, R., Quantum theory cannot be extended. arXiv:1005.5173v1 (2010)

\bibitem{as09} Suarez A., Nonlocal ``realistic'' Leggett models can be considered refuted by the before-before experiment. \emph{Found. Phys.} \textbf{38}, 583-589 (2008); On Bell, Suarez-Scarani, and Leggett experiments. \emph{Found. Phys.} \textbf{39}, 156–159(2009)

\bibitem{vang} Scarani V. and Gisin N., Superluminal influences, hidden variables, and signaling. {\em Phys. Lett. A}, \textbf{295}, 167 (2002).

\bibitem{as08.1} Suarez A., Quantum randomness can be controlled by free will -a consequence of the before-before experiment.	arXiv:0804.0871v2 (2008)


\end{references}
\end{document}